\documentclass[twocolumn,english,showpacs,floatfix,amsmath,amssymb,prb,superscriptaddress]{revtex4-1}

\usepackage{graphicx}
\usepackage{graphics}
\usepackage{amsmath}
\usepackage{amsfonts}
\usepackage{amssymb}
\usepackage{epstopdf}
\usepackage{makeidx}
\usepackage{epsfig,times}
\usepackage{bm}
\usepackage[colorlinks=true,citecolor=blue,linkcolor=blue,linktocpage=true,pagebackref=false]{hyperref}
\usepackage{color,soul} 
\usepackage{float}
\usepackage[english]{babel}

\newcommand{\be}{\begin{equation}}
\newcommand{\ee}{  \end{equation}}
\newcommand{\ba}{\begin{eqnarray}}
\newcommand{\ea}{  \end{eqnarray}}

\begin{document}

\title{A networked voting rule for democratic representation}

\author{Alexis R. Hern\'andez}
\affiliation{Instituto de F\'{\i}sica,
Universidade Federal do Rio de Janeiro, Rio de Janeiro, Brazil}
\affiliation{Institute for Biocomputation and Physics of Complex Systems (BIFI), Universidad de Zaragoza, Zaragoza, Spain}
\author{Carlos Gracia-L\'azaro}
\affiliation{Institute for Biocomputation and Physics of Complex Systems (BIFI), Universidad de Zaragoza, Zaragoza, Spain}
\author{Edgardo Brigatti}
\affiliation{Instituto de F\'{\i}sica,
Universidade Federal do Rio de Janeiro, Rio de Janeiro, Brazil}
\author{Yamir Moreno}
\affiliation{Institute for Biocomputation and Physics of Complex Systems (BIFI), Universidad de Zaragoza, Zaragoza, Spain}
\affiliation{Department of Theoretical Physics, Faculty of Sciences, Universidad de Zaragoza, Zaragoza, Spain}
\affiliation{ISI Foundation, Turin, Italy} 
\date{\today}

\begin{abstract}

We introduce a general framework for exploring the problem of selecting a committee of representatives with the aim of studying a networked voting rule based on a decentralized large scale platform, which can assure a strong accountability of the elected. The results of our simulations suggest that this algorithm-based approach is able to obtain a high representativeness for relatively small committees, performing even better than a classical voting rule based on a closed list of candidates. We show that a general relation between committee size and representatives exists in the form of an inverse square root law and that the normalized committee  size approximately  scales with the inverse of the community size, allowing the scalability to very large populations. These findings are not strongly influenced by the different networks used to describe the individuals interactions, except for the presence of few individuals with very high connectivity which can have a marginal negative effect in the committee selection process.

\end{abstract}

\maketitle

\section{Introduction}

The selection of an exemplar group of representatives to make 
decisions on behalf of a larger community  
is a widespread and critical problem for human societies \cite{condorcet,classical}.
Examples can be found in the 
election  of a legislative assembly 
in indirect democracies,  
in the elections for the trade union, 
for supervisory or faculty board, 
for executive officers or non-governmental organization boards. The most widely used electoral systems
can be classified into one of the following groups: first-past-the-post, two-round
systems, proportional representation, ranked voting or in a mix of two or more of the
previous groups \cite{Gallaguer1991,Boix1999,Gallagher2005,Blais2008,Bormann2013}. In general, these
systems seek to strike a balance between representativeness and effectiveness. In
the vast majority of electoral systems, including fully
proportional representation systems \cite{classical}, representatives gain a power of
representation that is not completely proportional to the number of voters
they represent, but rather the result of a given granularity. Interestingly, this
problem is also relevant for artificial 
systems, such as software multiagent systems, i.e. in recommendation 
systems \cite{recommendation} and election-based mechanisms  
for distributing data over an overlay P2P network \cite{P2P}. 

In this work, we address the general problem of selecting a group of candidates
that best represents the voters.
We consider systems where each voter is allowed to vote for only one candidate
and the elected are the ones 
who obtain a better rank among their counterparts.
In particular, we focus on the case of  multi-winner elections, 
that is choosing  a collective body of a given size (a committee).
We model an idealized situation where 
voters are rational individuals,
which means that they make 
a decision to maximize their representation,
and they present a general  
knowledge of the candidates 
and direct access to them.

In classical elections, a fixed number of candidates participates 
and voters rank the candidates expressing their preferences. 
We introduce a new formal model, where the list of candidates is not fixed in advance, but
they emerge as a self-organized process controlled by the voting rules. 
Moreover, voters express not preferences, but opinions, which determine
their indications about who they would like to see as their representative.

Our model introduces new mechanisms which give a fundamental importance
to the accountability of the elected committee.
The connection between representatives and constituents is fundamental and 
it is the basis for accountability, allowing to check for incompetence and corruption. 
In classical voting systems, this link 
can be generally insured only for small size community.
In particular, for the national legislative assembly, it can be partially controlled 
by the small size of the electoral district. Our model introduces a radical difference for obtaining
an efficacious accountability. 
In fact, it takes into account individuals first-hand trust relationship as a key ingredient to determine the elected representatives. 
Votes are assigned on the basis of a self-declared confidence circle, 
which is a network of trusted individuals which can be implemented
on an on-line platform.

After having implemented this new voting rule, its effects are 
tested modeling the behavior of the selected committee.
The committee runs a series of ballots 
making choices about different issues. 
The quality of  the elected committee is numerically assessed 
based on how much their final decisions are consistent 
with the personal opinions of the community.
Note that many works study scenarios
where representatives make decisions 
which are compared with some objective 
truth \cite{condorcet,truth}. In contrast, as in \cite{measureR}, our approach is interested in discriminating 
the selected boards which best 
represent the community opinions.


\section{The model}

The system is composed of a population of $N_e$ electors and an internet-based platform.
The platform allows the voters to self-declare who belongs to their confidence circle, which 
is a network of trusted individuals. 
The same platform is used by voters to manifest their 
opinions on $N_i$ issues. 
Issues are organized in questions which can be defined 
by a committee or by means of 
a self-organized process internal to the community. 
The answers of each individual $j$ are organized in a vector $v^j$. 
The vector is composed by $N_i$ cells and each cell can 
assume the value $1$ if the answer is positive, $-1$ if it is negative 
or $0$ if the question is left unanswered.

The following step allows to find
the better representative for  each confidence circle.
First, we consider an individual $j$ and we compute the vectors overlaps with all his neighbors $k$. This is obtained using the expression:
\begin{equation}
v^j*v^k=\frac{\sum_{m=1}^{N_i} (v^j_m \cdot v^k_m)\delta(v^j_m, v^k_m)}{\sum_{m=1}^{N_i}(v^j_m \cdot v^k_m)^2}\,\, ,
\label{eq:overlap_Demo}
\end{equation}
where the numerator counts the number of questions answered in the same way (only yes or not) and the denominator counts the number of questions answered simultaneously by both individuals. 
Each individual $j$ will indicate as his representative the individual $k'$
for which $v^j*v^{k'}$ is maximum. 
In the case where more than one individual generates the same maximum overlap value,
the individual with a higher connectivity is chosen as the representative. 
For the exceptional case when also the connectivity is equal, 
the representative is randomly selected between the 
similar ones.

\begin{figure}[h]
\begin{center}
\includegraphics[width=\columnwidth, angle=0]{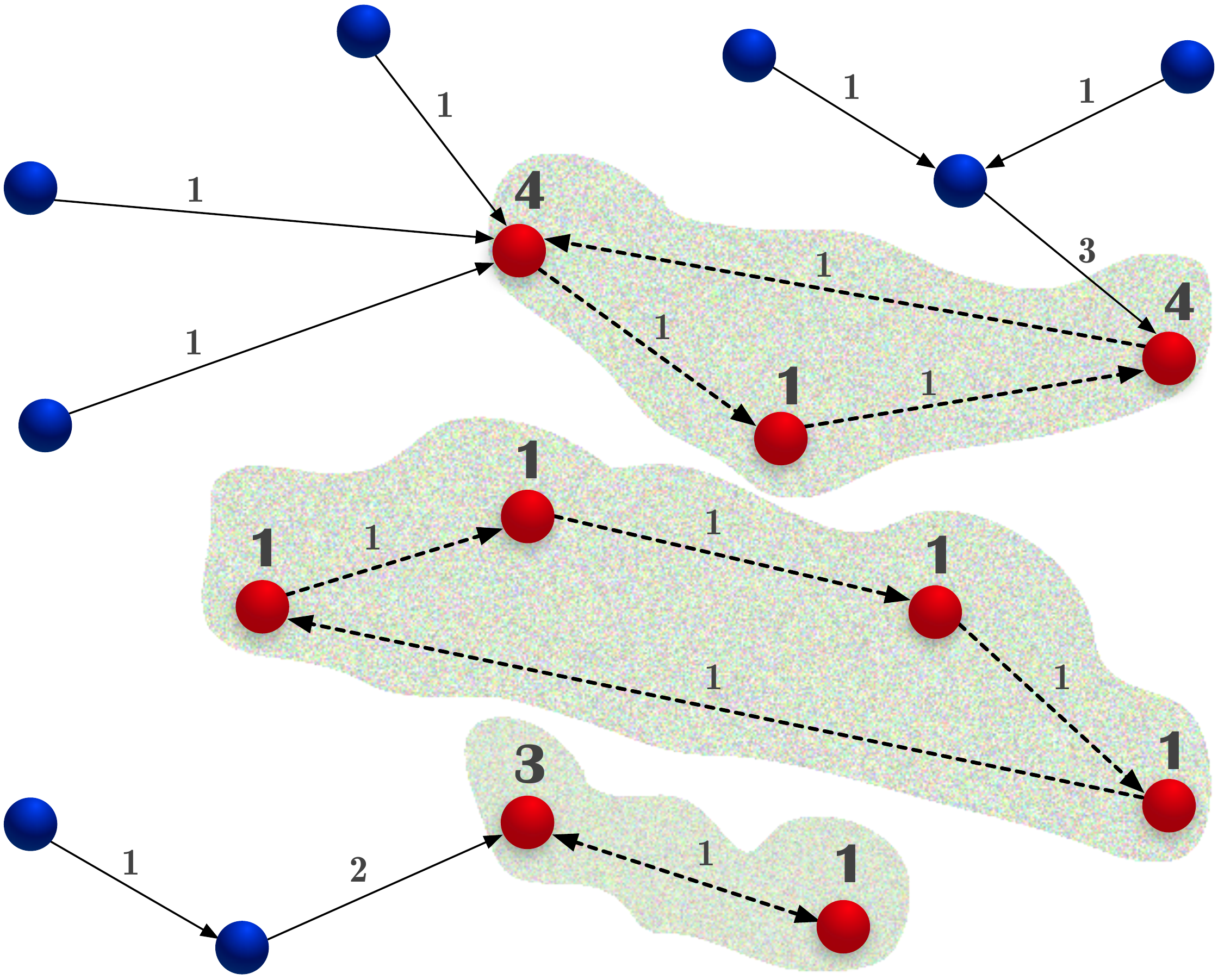}
\end{center}
\caption{Schematic representation of the vote process. 
Nodes stand for the individuals; the red ones belong to a cycle and will be confirmed as representatives if they collect more votes that the established threshold. 
The big numbers associated to the nodes represent the received cumulated votes. 
Arrows stand for the indication of each individual and the small numbers
associated to them represent the number of transferred votes.  
Dotted arrows belong to a cycle, where there is no cumulative transfer of votes.}
\label{fig_Vote_Flux}
\end{figure}

After the selection of the representative $k'$ for every voter $j$, as constrained by his
confidence network, 
the final step consists of choosing the aggregate of representatives of the entire community. 
To this end, we  construct a directed graph where a
node represents each individual and a directed link connects the individual  
with his personal representative. 
In this graph, which in general can be composed by different disconnected clusters, 
cycles are present. 
They represent individuals that have been mutually indicated by themselves.  
As all the individuals outside the cycles are represented by the individuals belonging to them, 
individuals who belong to cycles are the proper potential representatives for 
the community (see figure \ref{fig_Vote_Flux}). 

As a final step, 
among the individuals belonging to a cycle, only the ones with a number of votes 
larger than a threshold $\theta$ are indicated as representatives. 
Votes are counted considering the cumulative flow defined by 
the directed graph. If the individual $j$ is pointing to $z$, $z$ receives 
all the votes previously received by $j$ plus one.
This flow of votes is computed only following links outside the cycles.
Inside the cycles, only the single vote of an individual is counted.
In this way the number of representatives is reduced and results to be a fraction
of the total individuals which belong to a cycle.

\section{Results}

\subsection{General analysis}

The individuals' opinions in relation to the selected issues 
are randomly generated 
with the following rule: given an issue $i$,
an individual does not present an opinion ($v_i= 0$) with probability $1/3$. 
The probability to have an opinion $v_i=+1(-1)$, is $1/3+\epsilon_i$ $(1/3-\epsilon_i)$, 
where $\epsilon_i$ is a random variable following 
a normal distribution with mean value equal to zero 
and $\sigma^2=0.05$. 
Such a simplified 
characterization of the single opinions presents some contact with
real political opinions which, frequently, 
are polarized, presenting a natural bimodality of preferences 
in political and economical issues  \cite{dixit}.

The confidence circle of each individual is modelled 
generating a network where nodes represent individuals and links 
the trust relationships present in the community.
The confidence circle of an individual is obtained
selecting a node and considering its first neighbors.
Note that an important simplification of this approach 
is the fact that it generates individuals with
symmetric trust relationships. 
In the following analysis three types of networks are considered.
Homogeneous random networks, 
implementing the Erd\"os-R\'enyi model \cite{erdos}, where
the degree distribution is peaked around a typical value $\langle k\rangle$, 
heterogeneous networks,
using the Barabasi-Albert model \cite{barabasi},
with a power-law degree distribution $P(k) \propto k^{-3}$ and
networks with the small-world property using the  Watts and Strogatz model \cite{small}.
Our aim is not to model  specific aspects
of a real social network, but to use
simple examples just to discuss the possible 
influence  of some 
relevant network properties (such as the heterogeneity in 
the degree distribution, the average degree and the small-world property), on the behavior of our model.

The exploration of the system behavior can be 
obtained considering two fundamental observables.
The first one is the normalized committee size 
which is measured as the 
ratio between the number 
of elected individuals ($E$) and 
the total number of individuals of the community:
$F_{rep}=E/N_e$.
The second one is the representativeness.
This is defined measuring the fraction 
of decisions expressed by the elected 
committee ($e_j$) which matches with the community
decisions ($c_j$) over all the considered $N_i$ issues:
$R=\frac{\sum_{j=1}^{N_i} \delta(e_j-c_j)}{N_i}$.
As the opinions expressed by our community is 
bimodal, it follows that $R\ge0.5$.

The decision expressed by the elected 
committee is obtained through a majority
vote where each representative's vote is weighted 
by the numbers of popular votes he received
in the election procedure.
The community decision is obtained by a direct process (plebiscite), 
where every individual votes in accordance with the opinion
expressed in his vector $v^j$. Note that if the individual
has no opinion on a particular issue, he abstains from voting.

\begin{figure}[!t]
\includegraphics[width=\columnwidth, angle=0]{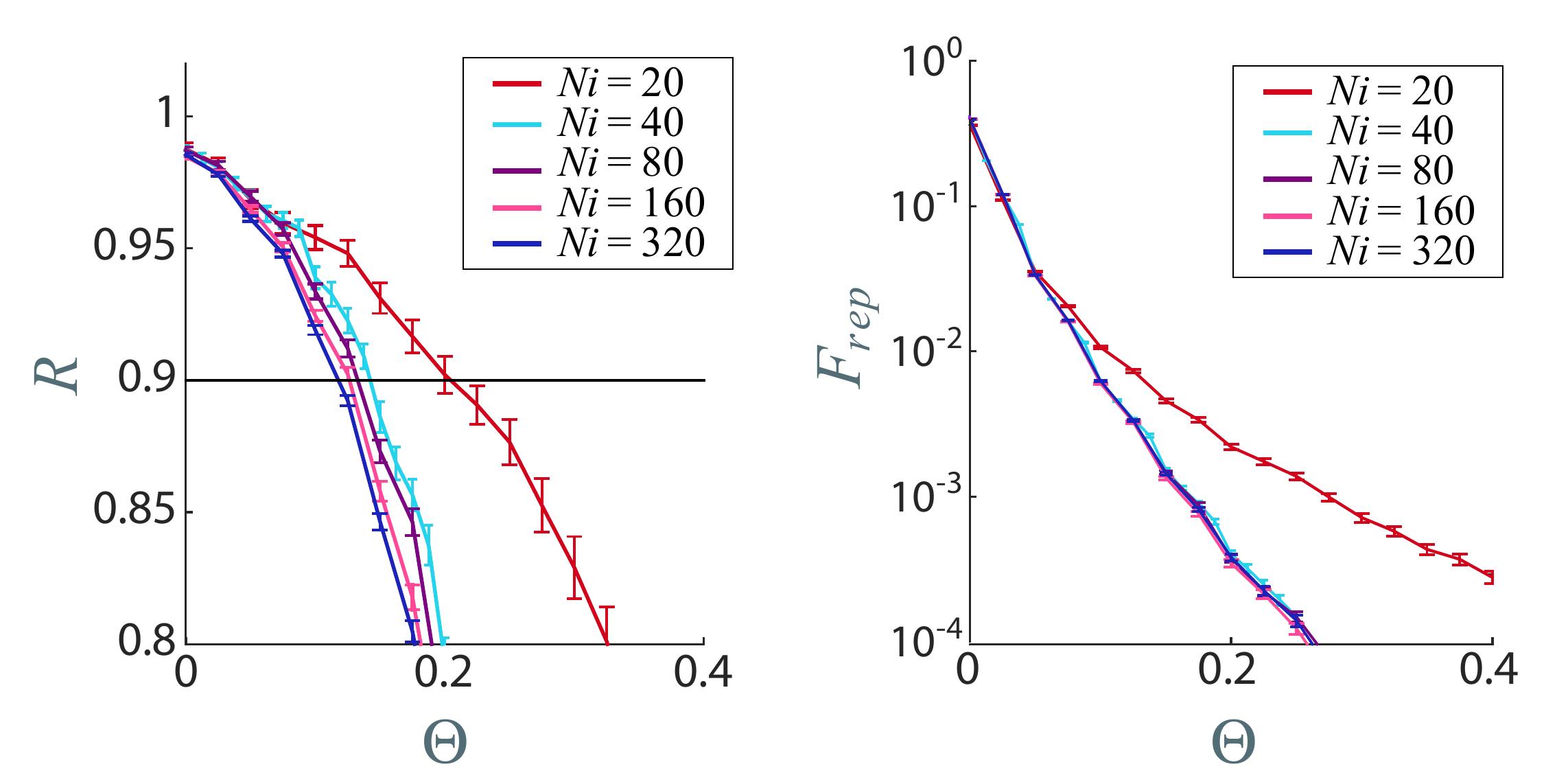}
\caption{Representativeness (left) and the  
normalized committee size (right, semi-logarithmic plot) as a function of the threshold $\Theta$ for different 
values of the number of issues $N_i$.
Both panels correspond to  $N_e=10000$ and confidence circles 
defined from a Erd\"os-R\'enyi network  with $\langle k \rangle =40$. 
Results are averaged over $100$ different realizations.}
\label{fig:Rep_vs_Thre}
\end{figure}

In Figure (\ref{fig:Rep_vs_Thre}) we show the representativeness $R$ and the 
normalized committee size $F_{rep}$
as a function of the threshold  $\Theta=\theta/N_e$ for different 
values of the number of issues $N_i$,
as obtained in a typical system with confidence circles 
defined from an Erd\"os-R\'enyi network. 
As expected, the representativeness and the
normalized committee size  decrease when increasing the 
threshold value. In particular,  the decrease of the normalized 
committee size is very fast.

For fixed threshold values,
decreasing the number of issues,
quite intuitively, increases the representativeness.  
Similarly, it increases the normalized committee size.
This last effect is not obvious and it 
is produced by the fact that
for a small number of issues 
the distribution of votes has a less
pronounced peak and the threshold
is not very efficient in selecting
between the indicated individuals. 
Anyway, this tendency rapidly saturates
and, for $N_{i} >40$, the curves do not show any 
relevant dependence on this parameter.

\begin{figure}[!t]
\includegraphics[width=\columnwidth, angle=0]{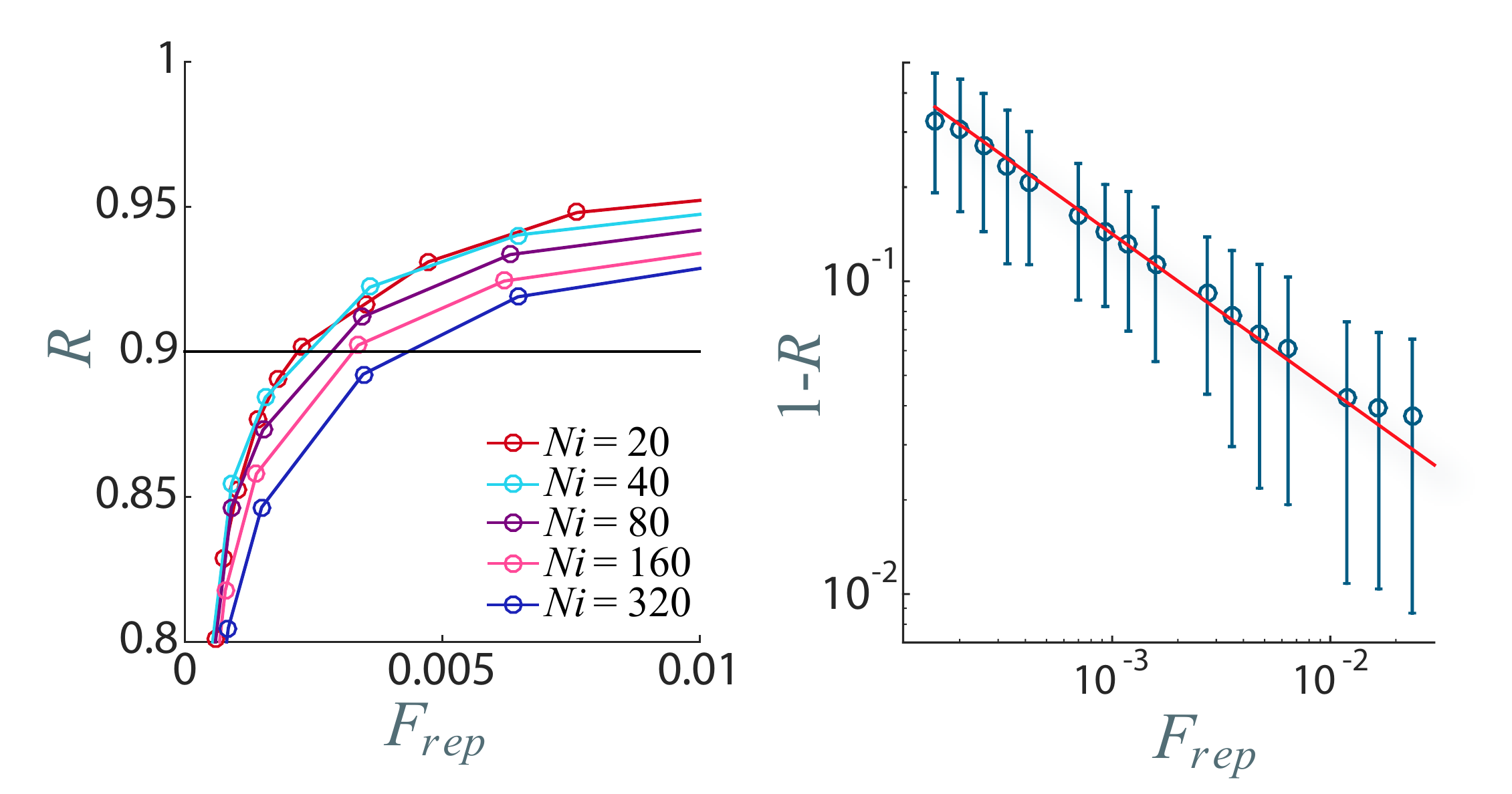}
\caption{ Left: Representativity vs normalized committee size 
Right: logarithmic plot of $1-R$ vs 
normalized committee size for $N_i=40$. The continuous line has slope $-1/2$. 
For both figures $N_e=10000$ and confidence circles 
are defined from a Erd\"os-R\'enyi network  with $\langle k \rangle =40$. 
Results are averaged over $100$ different realizations.}
\label{fig:Rel}
\end{figure}

The ideal committee corresponds to a small group of representatives 
which expresses a high level of representativeness. 
This is obtained selecting an intermediate value for $\Theta$, 
which can be identified seeking for a representativeness close to $0.9$,
and looking at the corresponding committee size.
For this reason, in the following we will plot 
the representativeness vs. the normalized committee size,
which allows a clear visualization of this fundamental relationship.
In Figure \ref{fig:Rel} 
we can observe that for fixed values of $R$, the normalized 
committee size increases when the number of issues increases. 

The relation between $R$ and $F_{rep}$,
 as reported in Figure \ref{fig:Rel}, can be characterized by a simple relation:
$1-R \propto 1/\sqrt{F_{rep}}$. 
of the obtained representativeness in relation to the ideal one 
scales as the square of  the inverse of the normalized committee size. 
This means that, for example,
for improving $R-1$ by a factor of 2, 
the number of elected  must quadruplicates.
This relation can be justified considering that $1-R$
is proportional to the error in the estimation of 
the mean opinion $v_i$ (the result of the plebiscite)
using a sample of size $F_{rep}$.
In the case where the estimation
of the mean opinion
uses an independent and identically distributed sample of size $n$,
it is well known that the standard error of the sample mean scales as 
 $n^{-1/2}$.  Interestingly, the same scaling law is preserved 
using our voting rule, which effectively 
can be seen as a particular data sample strategy.

\begin{figure}[!t]
\includegraphics[width=0.48\columnwidth, angle=0]{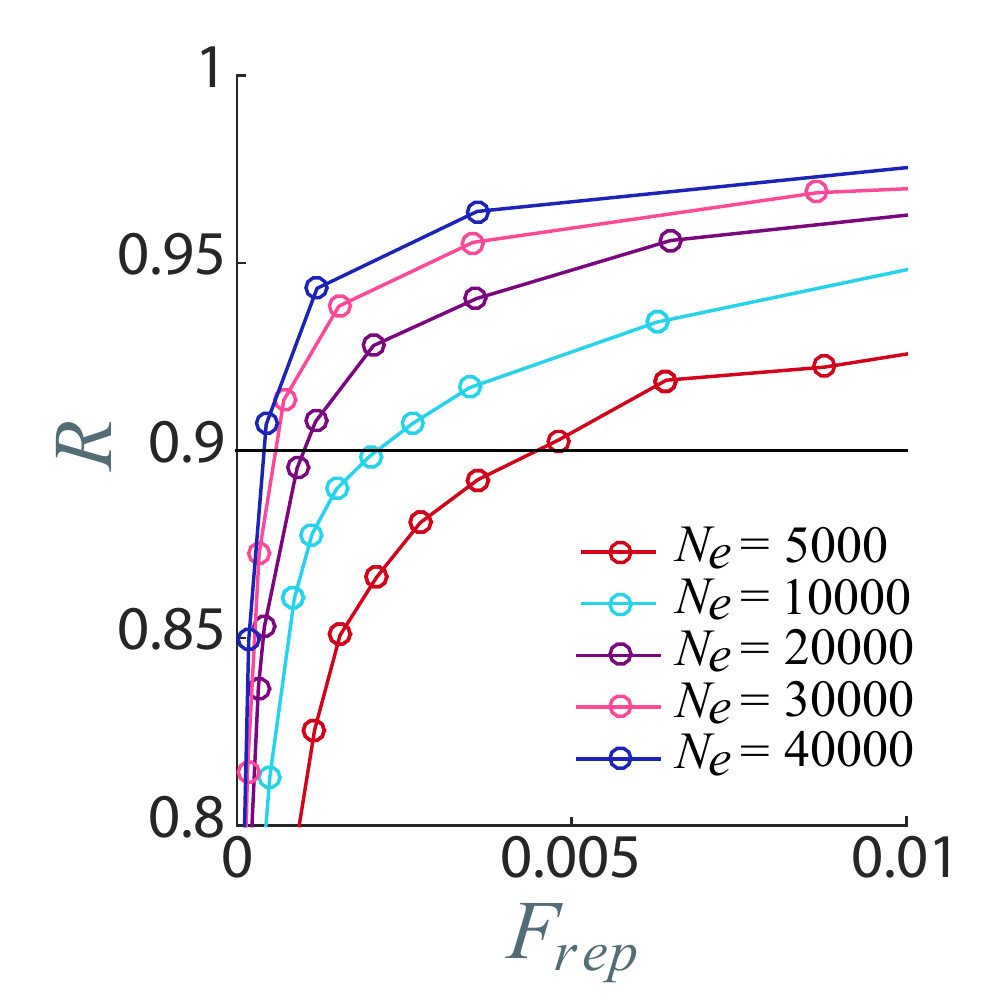} 
\includegraphics[width=0.48\columnwidth, angle=0]{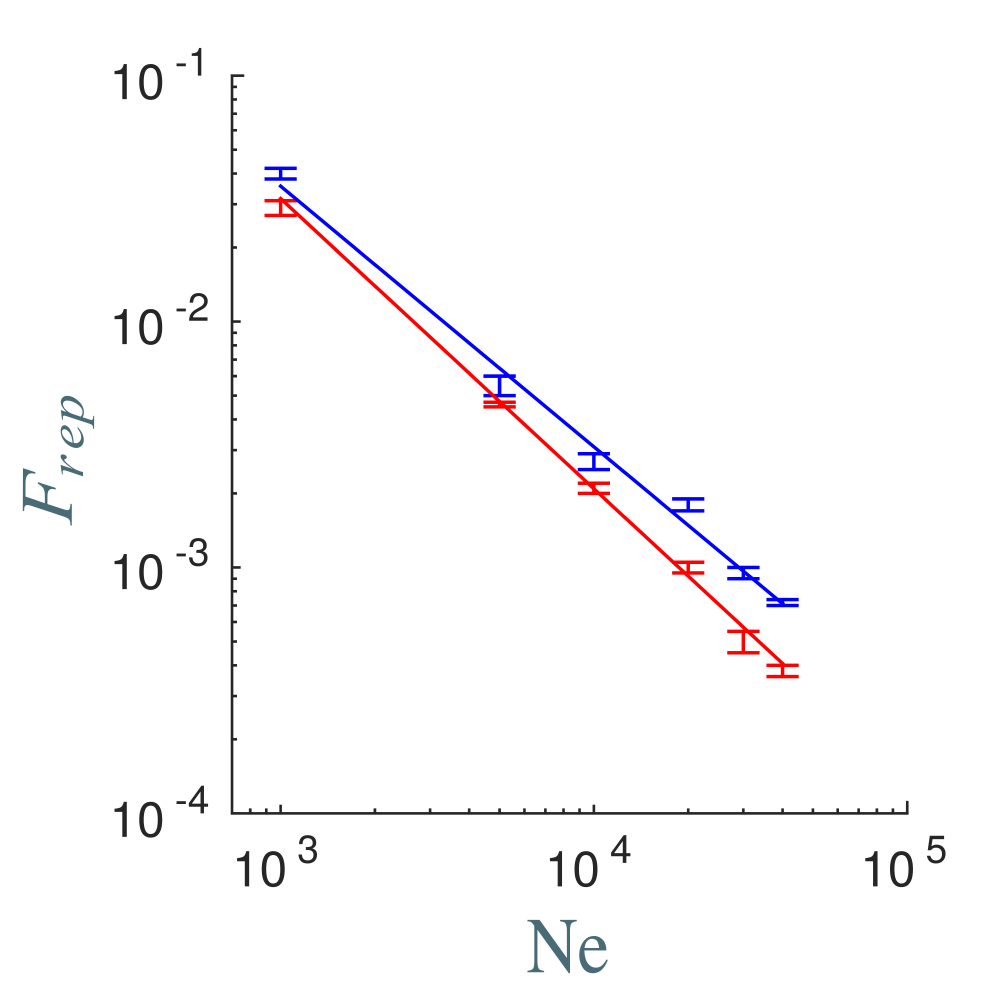}
\caption{ Left: representativity vs normalized committee size for a Erd\"os-R\'enyi network. 
Right: logarithmic plot of the normalized committee size fixing $R=0.9$, 
as a function of the number of electors.
Data are well approximated by a power-law fitting (continuous lines).
For the Barabasi-Albert network (blue) the power-law exponent is $-1.06\pm0.05$,
for the Erd\"os-R\'enyi network (red) $-1.18\pm0.03$.
The parameters used in the simulations are $\langle k \rangle =40$ and $N_i=40$ . 
Results are averaged over $100$ different realizations.}
\label{fig:Rep_vs_Frac_Ne}
\end{figure}

A similar analysis was conduced looking at the dependence on the 
system size $N_e$ (see Figure \ref{fig:Rep_vs_Frac_Ne}). 
Surprisingly, fixing $R$, the committee size decreases 
with the system size. For example, for the parameters
used in Figure \ref{fig:Rep_vs_Frac_Ne},
a representativity
of 0.9 is obtained with a committee of 29 members for a community of 1000 individuals, and with just 15 representatives for $N_e=30000$.
In particular, fixing $R=0.9$, 
$F_{rep}$ decreases approximately with the inverse
of $N_e$ using a Barabasi-Albert network, and even 
faster for an Erd\"os-R\'enyi network. 

\begin{figure}[!t]
\includegraphics[width=\columnwidth, angle=0]{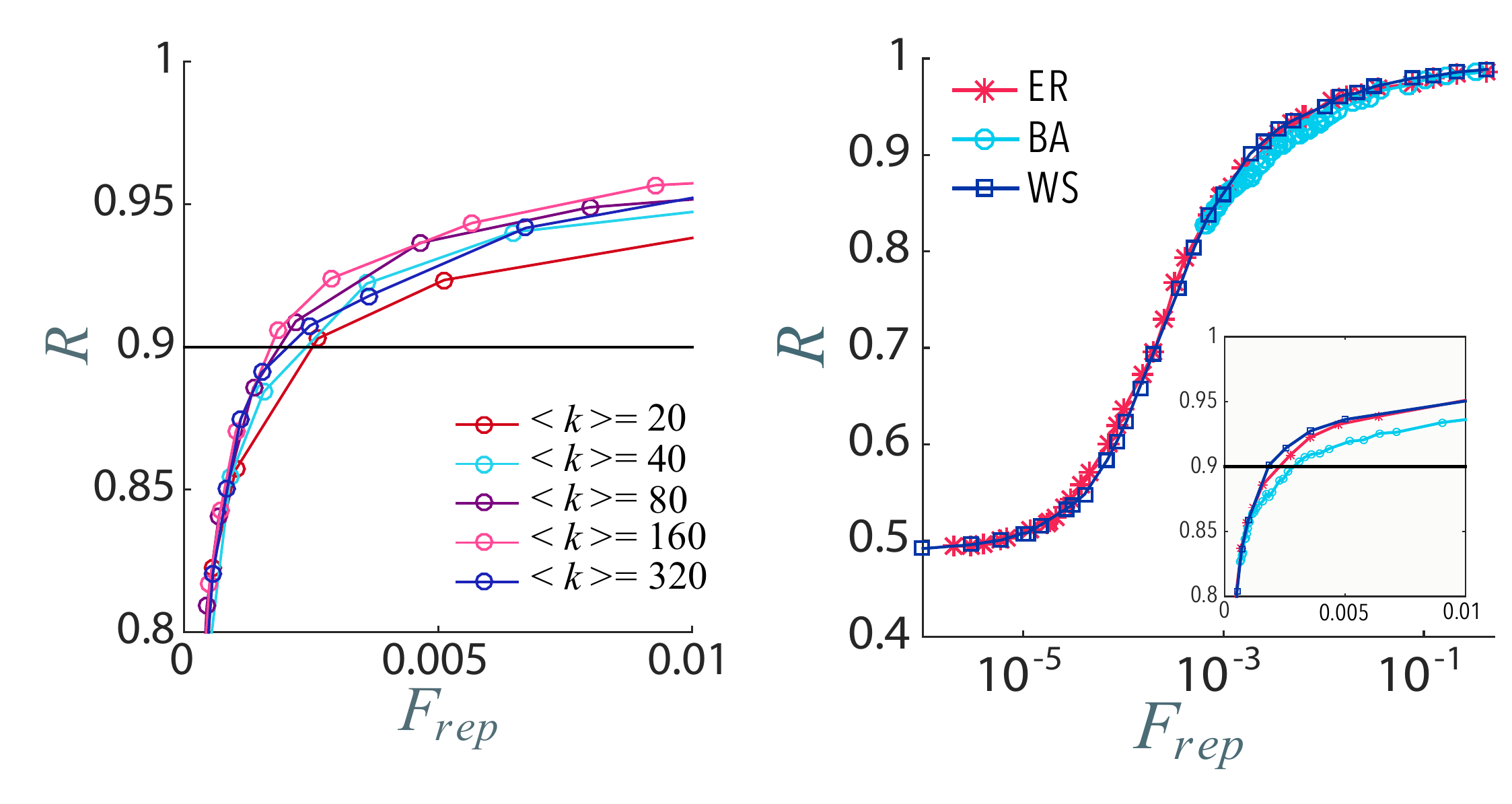}
\caption{ Representativity vs the fraction of elected citizens. On the left we consider a Erd\"os-R\'enyi network with different connectivities,  $N_e=10000$ and $N_i=40$. On the right we display the semi-logarithmic plot of three different networks: Erd\"os-R\'enyi, Barabasi-Albert and Watts-Strogatz, with $N_e=10000$, $N_i=40$, $\langle k \rangle = 40$. The Watts-Strogatz network has $\beta=0.1$. Results are averaged over $100$ different realizations.}
\label{fig:Rep_vs_Frac_K}
\end{figure}

In Fig. \ref{fig:Rep_vs_Frac_K} we can see that the representativeness is 
not strongly dependent on the connectivity of the network, and for 
$\langle k \rangle > 40 $ the curves 
present very similar behaviors. 
The heterogeneity in the degree distribution 
of the network seems to have a relative small impact on the results too, as it can be appreciated by comparing the results of the Erd\"os-R\'enyi network with the Barabasi-Albert one. 
Finally, the small world property of the Watts-Strogatz model does not influence our results.

\begin{figure}[!t]
\includegraphics[width=\columnwidth, angle=0]{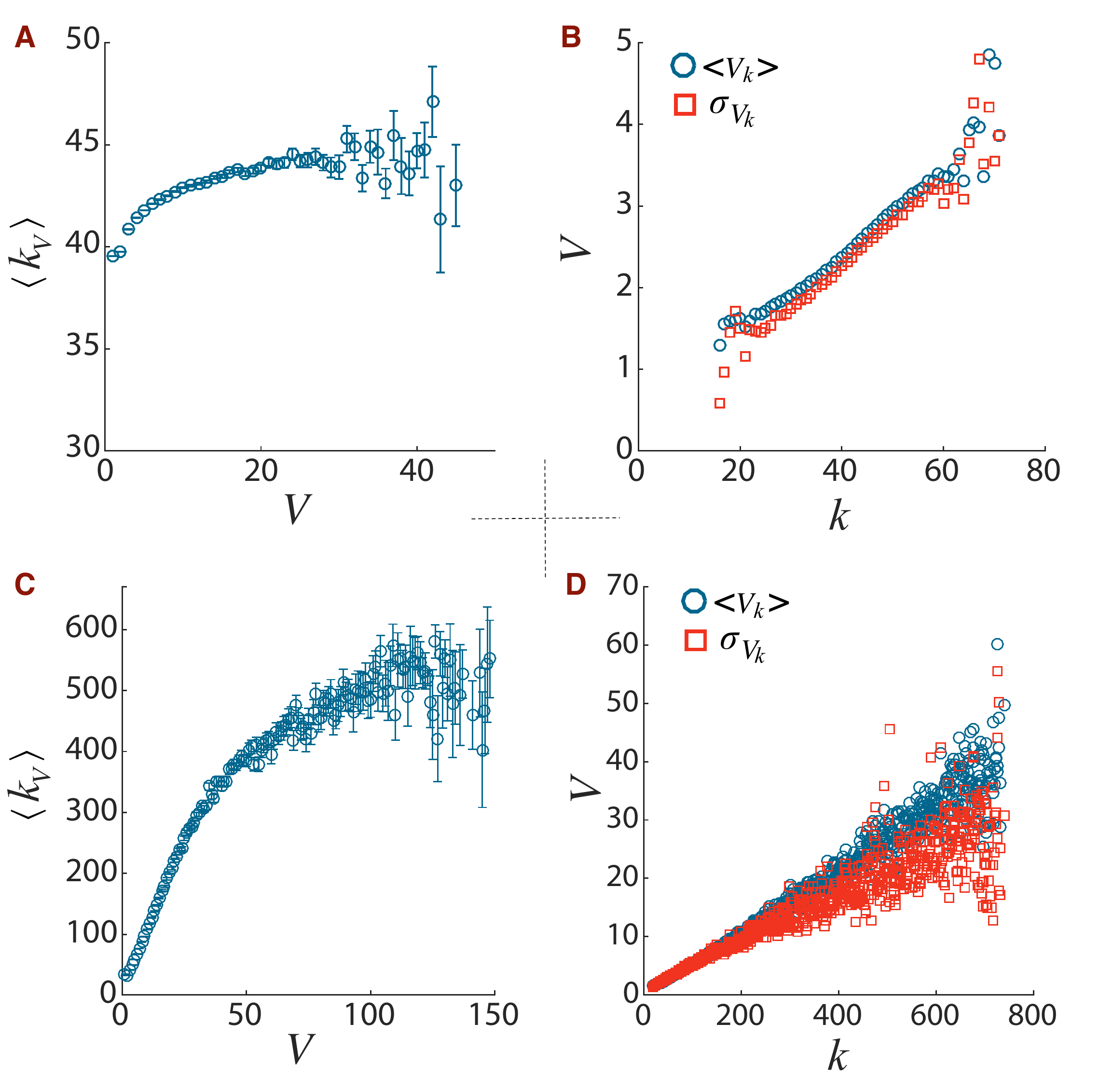}
\caption{ Left:  mean connectivity $\langle k_{_V} \rangle$ of a individual which received  a number of votes $V$. 
Right:  mean number of votes $\langle V_k \rangle$ collected by individuals with a given connectivity $k$. Data are obtained using $1000$ different realizations in a Erd\"os-R\'enyi network (top) 
and a Barabasi-Albert one (bottom), with $\langle k \rangle =40$, $N_i=40$ and $N_e= 10000$.}
\label{fig:Correlations}
\end{figure}

These results are consistent with the inspection of the relation between the individuals connectivity
and the number of votes they received.
As can be appreciated in Figure (\ref{fig:Correlations}), even if 
it is necessary a reasonable connectivity (higher than the mean value)
to obtain votes and a higher connectivity increases the probability to obtain more votes, 
for  Erd\"os-R\'enyi networks this effect is weak and not impactful. 
For the case of the Barabasi-Albert network, this effect is more relevant and probably it is the responsible
for the influence that this topology has on the behavior of the representativity and normalized committee size
(Fig. \ref{fig:Rep_vs_Frac_K}), determining a slightly weaker performance. 
In fact, it moderately decreases the representativeness for a fixed committee size, 
as higher connectivity generates a bias in the selection of the more representative individuals.

\subsection{Comparison with other voting rules}

In the following we explore if our model, at least 
theoretically, selects committees who
present a representativeness comparable with other traditional
methods of body selection. 
Obviously representativeness is compared among
committees of the same size.

\begin{figure}[!th]
\includegraphics[width=0.65\columnwidth, angle=0]{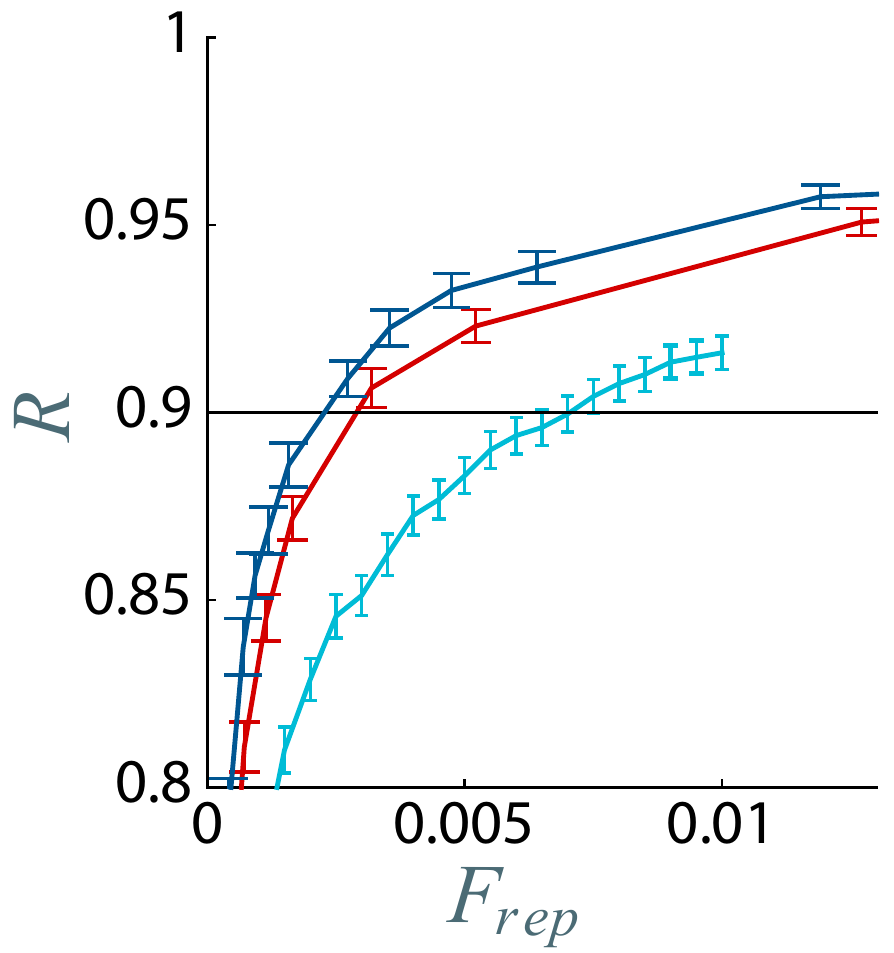}
\caption{Representativity vs the fraction of representatives for our model (blue curve), a model where 
a closed list of randomly selected politicians collect votes maximizing the overlap with the electorate (green curve) and a perfect voting rule (red curve).
Parameters: $N_e=10000$, $N_i=40$ and $\langle k \rangle=40$.  
}
\label{fig:Rep_vs_Frac_Comp}
\end{figure}

A first possible comparison is with a model of a  
traditional majority voting for the selection of representatives 
in a closed list of previously selected candidates. 
This is probably the most common practice in selecting committees. 
An example can be found 
in the election  of legislative assembly with the system of 
multi-member districts.
In our simulations, a list of $m$ candidates is randomly selected
among the community and each individual votes for the candidate
who presents the higher overlap with its opinion vector. Decisions are taken with the same weighted voting rule. 
This modeling approach mimics a voter
who presents a perfect knowledge of the candidates, 
and it assumes that he makes a rational decision to maximize his representation.
Also for this voting rule, representativeness is computed by comparing the decisions taken by the committee, obtained  with a weighted majority voting process, with the results of the direct popular vote.
 As can be appreciated in Figure \ref{fig:Rep_vs_Frac_Comp}, our model is by far more efficient, reducing the size of the committees in more than a half. Notice that if the voting process is not weighted, the difference is even bigger.\\

Finally, we compare our method to an idealized perfect
voting rule. This rule represents a situation of rational individuals
that have a perfect knowledge of all the 
other individuals, which means that they perfectly know the opinion of  all the other individuals.
Moreover they are globally networked, which means that 
they have a direct access to all other individuals, allowing to check their acts.
In this situation, a voter indicates an individual which presents the higher 
overlap with his opinion vector.
The selected committee is composed by the first $F_{rep}\cdot N_e$ individuals 
which poll more. 
Also in this case, the committee decisions are taken by means of a weighted majority vote.
This voting rule, although unrealistic, is still useful, at least, in two respects. 
Firstly, very small communities  can exhibit similar characteristics. Secondly,  the model is a useful yardstick for evaluating the levels of representativeness of other more realistic models.\\

Note that this voting rules can be considered, from a more abstract point of view, 
as an optimization problem: to find the vectors with greater overlap
in an ensemble of given vectors, with no other constraints.
In contrast, our original voting rule corresponds to the 
same optimization problem constrained by the fact that the overlap is inspected
only locally, on a small subset of the ensemble of given vectors, 
because of the presence of the confidence circles.

In Figure \ref{fig:Rep_vs_Frac_Comp} we compare 
the representativeness of the perfect
voting rule with our networked rule for different committee sizes.
It is quite impressive that the representativeness 
of our voting rule is comparable with the perfect voting rule. 
Actually, the voting rule here presented performs
even slightly better than the perfect rule, probably
due to the effects of the cumulated voting in the
weights used for selecting the committee decisions.
This fact suggests that the constraints introduced by the confidence
circles in the optimization of the opinions overlap are not effective 
in limiting the process of selecting the best overlaps 
because of the positive effects generated by the specific voting rule. 

\section{Discussion}

We introduced a general framework for studying the problem
of selecting a committee of representatives to make 
decisions on behalf of a larger community.
In our approach, we do not probabilistically
study some properties of a particular 
voting rule, but we realistically implement it
taking into account the effects of the individuals' opinions 
and of their social relations. 
Moreover, we 
model the behavior of the selected committee
which is called to make choices about different issues. 
In this way, we are able, not simply to
compare their choices
with some abstract truth, but
to clearly quantify the relation between
representativeness and committee size.

Based on this scheme, it is possible 
to study the properties of a new networked
voting rule, introduced with the aim of
obtaining high representativeness 
together with strong accountability 
of the elected.
This rule also presents the interesting feature that the
candidates are not fixed in advance, but they emerge as 
a self-organized process. 

The results of our simulations
suggest that this rule 
is able to obtain a 
high representativeness 
for relatively small committees.
In fact, these outputs are comparable 
with an ideal perfect voting rule,
and they perform clearly better
than a classical voting rule based 
on a closed list of candidates.
Moreover, for fixed representativeness, the normalized committee  size 
approximately  scales with the inverse of the community size,
allowing the scalability of this approach to very large populations.
Finally, we were able 
to characterize the relation between 
the committee size and the representatives
by means of a general inverse 
square root law.\\

These findings are robust
and they are not strongly influenced 
by general properties of 
the network used to describe the 
individuals interactions.
It seems that only the heterogeneity 
can have some role in modifying the relation
between representativeness and
committee size. In fact, the presence of 
a few individuals with very high connectivity
can negatively influence the selection
process of the committee.
This fact is consistent with the
risks related to a dominant position of single individuals, 
which may lead to a general sub-representation of 
the general opinion of the community.\\

The introduction of this networked 
voting rule, based on a decentralized large scale platform, can be considered 
as an interesting tool for implementing an hyper-representative 
mechanism of committee selection 
based on  a distributed social mechanism
where the use of a block chain encryption mechanism
could guarantee the security of the voting process. 
This algorithm-based approach facilitates the participation of the entire population both as electors and as representatives, dismissing the importance of traditional authorities.
Our analysis highlights the feasibility and potential of this election rule 
from the point of view of an organizational theory; specific and thorny 
political theory considerations and risks are not addressed.

Data accessibility: The code used for this research is available at \href{https://www.researchgate.net/publication/321804236_AxelDemX3}{code}

Competing interests: We have no competing interests.

Author contributions: A. R. H. is responsible for the conceptual idea behind the paper; 
All the authors participated in the design of the model;
A. R. H. and C. G. L. have written the code and ran the simulations; 
All authors participate in the analysis and discussions of the results; 
E. B. was the main responsible in the writing process but all authors 
commented on the manuscript and gave final approval for publication.

Acknowledgements: A. R. H. thanks COSNET Lab at the Institute BIFI for partial support and hospitality during the realization of most of this work.

Funding statement: C. G. L. and Y. M. acknowledge support from the Government of Arag\'on, Spain through a grant to the group FENOL, by MINECO and FEDER funds (grant FIS2014-55867-P) and by the European Commission FET-Open Project Ibsen (grant 662725).


\begin{thebibliography}{99}

\bibitem{condorcet}
Condorcet MJ. 1785 Essai sur l'application de l'analyse à la probabilité des décisions rendues à la pluralité des voix (p. ). de l'Imprimerie royale. https://doi.org/10.3931/e-rara-3791

\bibitem{classical}
Kelly J 1991 Social Choice Bibliography. Social Choice and Welfare, 8(2), 97-169. Retrieved from http://www.jstor.org/stable/41105976
Chamberlin J, Courant P 1983 Representative Deliberations and Representative Decisions: Proportional Representation and the Borda Rule. The American Political Science Review, 77(3), 718-733. doi:10.2307/1957270
Monroe B 1995 Fully Proportional Representation. The American Political Science Review, 89(4), 925-940. doi:10.2307/2082518

\bibitem{Gallaguer1991} 
Gallagher M 1991 Proportionality, disproportionality and electoral systems. Electoral Studies, 10(1), 33-51. https://doi.org/10.1016/0261-3794(91)90004-c

\bibitem{Boix1999} 
Boix C 1999 Setting the Rules of the Game: The Choice of Electoral Systems in Advanced Democracies. The American Political Science Review, 93(3), 609-624. doi:10.2307/2585577

\bibitem{Gallagher2005}
Gallagher M, Mitchell P, (Eds.) 2005 The Politics of Electoral Systems. Oxford University Press. https://doi.org/10.1093/0199257566.001.0001

\bibitem{Blais2008} 
Blais A. (2008, May 8). To Keep or To Change First Past The Post? Oxford University Press. https://doi.org/10.1093/acprof:oso/9780199539390.001.0001

\bibitem{Bormann2013}
Bormann NC, Golder M 2013 Democratic Electoral Systems around the world, 1946-2011. Electoral Studies, 32(2), 360-369. https://doi.org/10.1016/j.electstud.2013.01.005

\bibitem{recommendation} 
Lu T, Boutilier C 2011 Budgeted social choice: from consensus to personalized decision making, Proceedings of IJCAI-2011, 280, DOI:
https://doi.org/10.5591/978-1-57735-516-8/IJCAI11-057
Naamani-Dery L, Kalech M, Rokach L, Shapira B 2014 Preference elicitation for narrowing the recommended list for groups. In Proceedings of the 8th ACM Conference on Recommender systems - RecSys-14. ACM Press. https://doi.org/10.1145/2645710.2645760

\bibitem{P2P} 
Malpani N, Welch JL, Vaidya N 2000 Leader election algorithms for mobile ad hoc networks. In Proceedings of the 4th international workshop on Discrete algorithms and methods for mobile computing and communications - DIALM-00. ACM Press. https://doi.org/10.1145/345848.345871
Ramanathan M K, Ferreira RA, Jagannathan S, Grama A, Szpankowski W 2007 Randomized leader election. Distributed Computing, 19(5-6), 403-418. https://doi.org/10.1007/s00446-007-0022-4
Baraglia R, Dazzi P, Mordacchini M, Ricci L, Alessi L 2011 On Democracy in Peer-to-Peer systems arXiv:1106.3172v1

\bibitem{truth}
Feddersen T, Pesendorfer W 1998 Convicting the Innocent: The Inferiority of Unanimous Jury Verdicts under Strategic Voting. The American Political Science Review, 92(1), 23-35. doi:10.2307/2585926
Myerson RB 1998 Extended Poisson Games and the Condorcet Jury Theorem. Games and Economic Behavior, 25(1), 111-131. https://doi.org/10.1006/game.1997.0610
Young P 1995 Optimal Voting Rules. The Journal of Economic Perspectives, 9(1), 51-64. Retrieved from http://www.jstor.org/stable/2138354

\bibitem{measureR} 
Skowron P 2015 What Do We Elect Committees For? A Voting Committee Model for Multi-Winner Rules, Proceedings of the Twenty-Fourth International Joint Conference on Artificial Intelligence, 1141 (2015). http://dl.acm.org/citation.cfm?id=2832249.2832407

\bibitem{dixit} 
Dixit AK, Weibull JW 2007 Political polarization. Proceedings of the National Academy of Sciences, 104(18), 7351-7356. https://doi.org/10.1073/pnas.0702071104
Gravino P, Caminiti S, Sîrbu A, Tria F, Servedio VDP, Loreto V 2016 Unveiling Political Opinion Structures with a Web-experiment. In Proceedings of the 1st International Conference on Complex Information Systems. SCITEPRESS - Science and and Technology Publications. https://doi.org/10.5220/0005906300390047

\bibitem{erdos} 
Erd\"os P, R\'enyi A 1959 On random graphs, I, Publ. Math. (Debrecen) {\bf 6}, 290.
\bibitem{barabasi} 
Barab\'asi AL, Albert R 1999 Emergence of Scaling in Random Networks. Science, 286(5439), 509-512. https://doi.org/10.1126/science.286.5439.509
\bibitem{small} 
Watts DJ, Strogatz S H 1998 Collective dynamics of "small-world" networks. Nature, 393(6684), 440-442. https://doi.org/10.1038/30918

\end{thebibliography}
\end{document}